\documentclass{ieeeaccess}
\usepackage{cite}
\usepackage{amsmath,amssymb,amsfonts}
\usepackage{algorithmic}
\usepackage{graphicx}
\usepackage{textcomp}
\usepackage{xurl}
\usepackage{hyperref}

\usepackage{bm}
\makeatletter
\AtBeginDocument{\DeclareMathVersion{bold}
\SetSymbolFont{operators}{bold}{T1}{times}{b}{n}
\SetSymbolFont{NewLetters}{bold}{T1}{times}{b}{it}
\SetMathAlphabet{\mathrm}{bold}{T1}{times}{b}{n}
\SetMathAlphabet{\mathit}{bold}{T1}{times}{b}{it}
\SetMathAlphabet{\mathbf}{bold}{T1}{times}{b}{n}
\SetMathAlphabet{\mathtt}{bold}{OT1}{pcr}{b}{n}
\SetSymbolFont{symbols}{bold}{OMS}{cmsy}{b}{n}
\renewcommand\boldmath{\@nomath\boldmath\mathversion{bold}}}
\makeatother

\def\BibTeX{{\rm B\kern-.05em{\sc i\kern-.025em b}\kern-.08em
    T\kern-.1667em\lower.7ex\hbox{E}\kern-.125emX}}

\begin{document}
\history{Date of publication xxxx 00, 0000, date of current version xxxx 00, 0000.}
\doi{10.1109/ACCESS.2024.0429000}

\title{Towards a Definition of the Computational Architecture of Open Scholarly Infrastructures}
\author{\uppercase{Ivan Heibi}\authorrefmark{1,3}, 
\uppercase{Mario Petrella}\authorrefmark{1}, \uppercase{Angelo Di Iorio}\authorrefmark{2}, AND \uppercase{Silvio Peroni}\authorrefmark{1,3}
}

\address[1]{Research Centre for Open Scholarly Metadata, Department of Classical Philology and Italian Studies, University of Bologna, 40126 Bologna, Italy}
\address[2]{Department of Computer Science and Engineering, University of Bologna, 40126 Bologna, Italy}
\address[3]{Digital Humanities Advanced Research Centre, University of Bologna, 40126 Bologna, Italy}

\tfootnote{This work was funded by the European Union’s Horizon Europe programme, Grant Agreement No. 101188018 (GRAPHIA).}

\markboth
{Author \headeretal: Preparation of Papers for IEEE TRANSACTIONS and JOURNALS}
{Author \headeretal: Preparation of Papers for IEEE TRANSACTIONS and JOURNALS}

\corresp{Corresponding author: Ivan Heibi (e-mail: ivan.heibi2@unibo.it).}

\begin{abstract}
This paper proposes a layered framework for defining the architecture of the computational unit of an Open Scholarly Infrastructure (OSI) and presents a concrete implementation possibility through the introduction of OpenCitations, an OSI dedicated to publishing citation data and bibliographic metadata. Grounded in the Principles of Open Scholarly Infrastructure (POSI), the study focuses on the technical dimensions of openness, sustainability, interoperability, and reproducibility that are required for a robust OSI. The proposed approach first identifies the core principles and technical features that a computational block in an OSI should guarantee, including separation of operational domains, orchestration, scalability, observability, automation, and workflow portability. It then maps these requirements onto a layered architectural model composed of Hardware, Virtualization, Orchestration, and Application layers, complemented by a transversal Meta layer. The OpenCitations case study shows how this framework can be instantiated in practice through an on-premises infrastructure. The case is presented with details regarding the technicalities and actual implementations of the proposed methodology. By combining a conceptual definition with a real-world implementation, this work offers both a practical reference and a methodological basis for designing the computational block of an OSI. 
\end{abstract}


\titlepgskip=-21pt

\maketitle

\section{Introduction}
\label{sec:introduction}
\PARstart{T}{he} design and operation of Open Scholarly Infrastructures (OSIs) – shared research infrastructures, whether physical or virtual, including scientific equipment, scholarly resources, data repositories, research information systems, bibliometric platforms, computational services, and digital infrastructures that support open science across communities – have become central to the viability of open science and open research infrastructures \cite{b17,b2}. As scholarly communication moves toward greater openness and transparency, OSIs are expected to provide not only access to research outputs, but also reliable, interoperable, and long‑term services that support the reuse, assessment, and reproducibility of scholarly knowledge \cite{b7,b12}. 

In recent years, organizations such as SPARC (\url{https://sparcopen.org/}), the Invest in Open Infrastructures (IOI, \url{https://investinopen.org/})  and the Global Sustainability Coalition for Open Science Services (SCOSS) (\url{https://scoss.org/}) have increasingly recognized the strategic importance of OSIs and have begun to invest directly in supporting their governance, funding, and long‑term sustainability \cite{b18}. These initiatives signal a shift from a purely publisher‑centric model of scholarly communication toward infrastructures that are community‑owned, technically robust, and aligned with the principles of open science.  

Recently, also the CoARA Working Group on Open Infrastructures for Responsible Research Assessment (OI4RRA) provided a framework and conceptual architecture for implementing responsible research assessment through open, interconnected, and sustainable infrastructures. It outlines key principles, technical requirements, and governance practices, and proposes a model covering persistent identifiers and metadata standards, open publishing, metadata aggregation, and assessment support services. The work supports Organisations in transitioning from proprietary systems towards transparent, community-driven infrastructures \cite{b4,b5}.

Within this context, this article proposes a layered framework for defining the computational architecture of Open Scholarly Infrastructures and demonstrates its concrete implementation through the case of OpenCitations (\url{https://opencitations.net}), an OSI dedicated to publishing citation data and bibliographic metadata \cite{b6}. 

The work is grounded in the Principles of Open Scholarly Infrastructure (POSI) \cite{b7} – a set of guidelines that open scholarly infrastructure organisations and initiatives supporting the research community can adopt to assess and monitor updates and progress in open science practices by OSIs. Our study focuses on the part of POSI that addresses the technical (i.e. software) dimensions of openness, sustainability, interoperability, and reproducibility required for a robust computational component in the OSI. The proposed approach first identifies the core principles and technical features that an OSI should guarantee, including separation of operational domains, orchestration, scalability, observability, automation, and workflow portability. It then maps these requirements onto a layered architectural model composed of Hardware, Virtualization, Orchestration, and Application layers, complemented by a transversal Meta Layer for automation and reproducibility.  

The OpenCitations case study shows how this framework can be instantiated in practice through an on‑premises infrastructure. The analysis is grounded in the actual technical choices, service deployments, and orchestration practices adopted by the project, and it is presented with a practical real-case scenario of project management within OpenCitations, i.e., the GraspOS project (\url{https://graspos.eu/}).

The choice of OpenCitations as a case study is strongly motivated by the authors’ direct involvement in the infrastructure, as they are members of the OpenCitations team. In addition, OpenCitations’ implementation strategies and infrastructure design have recently been informed by the team’s participation in the Executive Master in Management of Research Infrastructures (EMMRI, \url{https://emmri.unimib.it/}). Through this programme, the team gained further knowledge and practical experience in managing research and open science infrastructures. That experience provided an important foundation for the development of this work.

The objective is not only to document a specific implementation, but to present a generalisable approach that can inform and support other OSIs. By abstracting from the OpenCitations experience, this work aims to highlight design principles and technical practices that may be adopted or extended by similar initiatives operating in different contexts.

The rest of this paper is organized as follows. Section \ref{sec:guidelines} introduces the guidelines that underpin our approach by discussing POSI, outlining the main operational units of an OSI, and identifying the key computational features that such an infrastructure should guarantee. Section \ref{sec:method} presents the proposed methodology through a layered architectural model, describing the role of each layer and its contribution to the definition of the OSI computational component. Section \ref{sec:oc} applies the methodology to the OpenCitations case study, first by analysing its computational architecture, then by illustrating an operational scenario based on the GraspOS project, and finally by discussing elements for evaluating the infrastructure. Section \ref{sec:conclusions} concludes the article, summarizing and discussing the main outcomes of the work and outlining directions for future developments.w

\section{Guidelines}
\label{sec:guidelines}
Before presenting the proposed methodology for designing an OSI, it is necessary to identify and articulate the foundational principles that such an infrastructure should adhere to. These principles are aligned with the Principles of Open Scholarly Infrastructure (POSI) \cite{b7}. Building on this basis, we define a comprehensive set of features that the technical infrastructure should incorporate. Each feature is then examined individually to determine the most appropriate strategy for its implementation and integration within the overall architectural design of the infrastructure.

In other words, this section outlines the fundamental guidelines underpinning the methodology introduced in the following section. To this end, the section first discusses the POSI and their relevance to this work, then provides an overview of OSIs and their main operational units, and finally sketches the set of features that should be taken into account in the definition of the OSI computational block.

\subsection{Principle of Open Scholarly Infrastructure (POSI)}
An OSI should consistently adhere to the Principles of Open Scholarly Infrastructure (POSI) \cite{b7}. POSI encompasses multiple infrastructure dimensions, each involving different operational units. Since the main focus of this work is strictly related to the computational and technical block of an OSI, this section will mainly focus on the subset of principles directly relevant to that part. In particular, we consider those principles that translate into concrete technical responsibilities, such as system architecture, service/data management or technical operability.

The primary guiding principle in the architectural design of an OSI is the \textbf{"Living Will"}, which requires defining and clearly communicating long-term stewardship commitments. These commitments should specify the technical procedures and the handling of data, resources, and services in the event of their transfer to a successor or the orderly wind-down of the organization or service. Coherently, the other principles should follow to ensure this behaviour.

\textbf{"Open source"}, \textbf{"Transparent operations"}, and \textbf{"Revenue generated from services, not data"} are essential in this regard. Open-source technologies and open data practices play a central role in developing an effective OSI \cite{b8}. These elements also have a significant positive impact on other key aspects that our proposed infrastructure should uphold, particularly reproducibility \cite{b8}. This principle should be taken into account both when the software and data are produced within the infrastructure itself and when they are supplied or taken by external third parties.

Working with open-source software provides transparency, allowing researchers to inspect, modify, and validate the underlying code that drives analytical workflows and data services. This openness not only strengthens trust in the infrastructure but also enables collective improvement and long-term sustainability through community-driven development. On the other hand, equally essential is the use of open data formats and standards, which ensure that research outputs remain accessible, interoperable, and reusable across disciplines and institutions. By prioritizing open metadata schemas, FAIR-compliant repositories \cite{b19}, and standardized APIs, the infrastructure facilitates seamless data exchange and supports reproducibility at scale. These principles, alongside the POSI guideline \textbf{"Prioritise interoperability and open standards"}, ensure continuity and resilience of the infrastructure.

The use of open-source technologies and openly available data is naturally linked to the purpose of the infrastructure and its governing policies. We recognize that, in many cases, these principles cannot be fully upheld. For example, certain copyright restrictions and/or other legal constraints (e.g. GDPR) may limit the use or public sharing of data. Likewise, some proprietary software or tools may be indispensable when they are the only viable solution for a specific task, such as designing complex 3D models. Given these constraints, the ideal approach is to remain as open as possible \cite{b20}. The more consistently this strategy is followed, the easier it becomes to ensure other important aspects, such as the overall reproducibility of the infrastructure, are achieved smoothly and effectively.

\subsection{OSI architecture}
\begin{figure}[t!]\centering\includegraphics[width=0.4\textwidth]{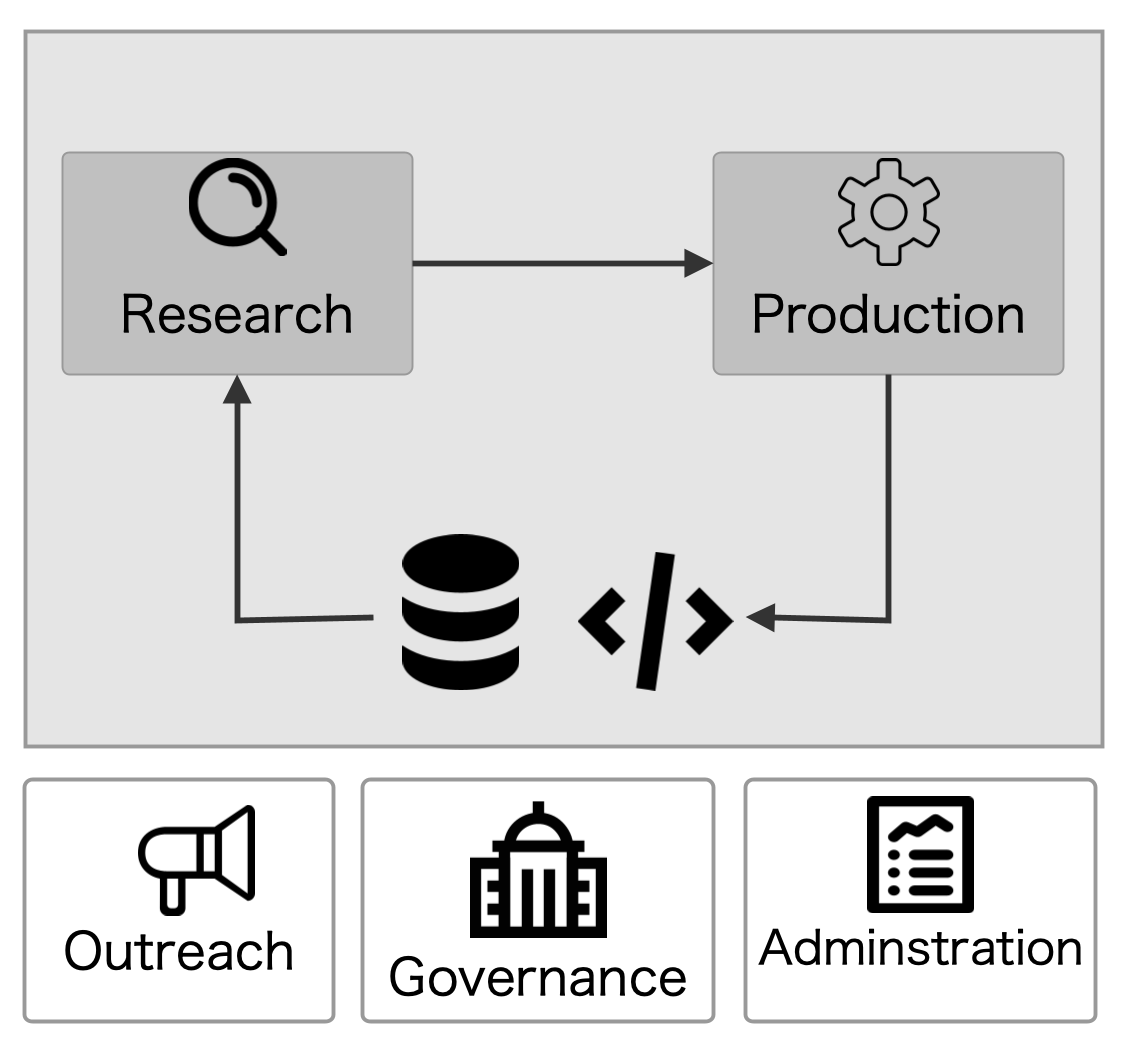}\caption{\textbf{OSI overview. The computational components (gray background) form a continuous cycle linking the Research, Production units with the produced Databases and Services. Other strategic units are also depicted (white background): Outreach, Governance, and Administration}\label{fig1}}\end{figure}

On Figure~\ref{fig1}, we illustrate a high-level abstract architecture of an OSI as we have conceptualized it. We highlight all the main units composing an OSI; the technical and computational block is represented with a gray background: this will be the core focus of our work. The proposed overview should be seen as a conceptual synthesis rather than a direct adoption of an existing reference model. It abstracts recurring functionalities of an OSI documented in the literature. Mainly from the documentation provided by the POSI \cite{b7}. In such documentation, governance, administration and operations, outreach and community engagement, and the technical production of data and services are clearly addressed.

Although our primary emphasis is on the technical and computational block of an OSI, a brief discussion of the other core units, which form the main essential components that contribute to the overall composition of the infrastructure, is useful and essential for the understanding of how the computational block works (in Figure~\ref{fig1} with a white background).

Looking at the components outside the computational block, the Governance unit plays an overarching role in defining policies, standards, and the strategic direction of the infrastructure. It ensures that activities across administration, research, production, and outreach are aligned with an established roadmap guiding both current operations and future developments. These operations should also follow the principles that an OSI should adhere to – i.e., the POSI. Governance is also responsible for high-level decision-making and the allocation of resources across the system. In contrast, while the Governance unit focuses on shaping strategic directions in both the short and long term, the Administration unit is responsible for their practical execution. Specifically, it oversees the operational implementation and enforcement of governance policies, including user management, resource allocation, service lifecycle management, and the day-to-day coordination of the infrastructure.

Finally, the Outreach unit is responsible for disseminating information about the OSI and engaging with its user community. It promotes awareness of available services, research outputs, and infrastructure developments through publications, announcements, and regular updates. The execution of outreach activities may, in some cases, rely on services provided by the computational block, such as web platforms supporting the infrastructure’s online presence.

The computational block of an OSI, as we intend for our OSIs, is organized as a continuous cycle connecting the units Research, Production (i.e. the technical implementation), and the database/services which are deployed.

The Research unit represents the core environment in which new software, tools, and datasets are first conceptualized, developed, and evaluated. The activities within this unit can be organized through multiple research projects and activities, each aligned with the strategic and technical directions defined at the governance level. These projects serve as the primary drivers for experimentation, prototyping, and methodological development within the infrastructure. Research decisions and outputs may align directly with specific project objectives. Alternatively, they can support standalone initiatives, such as independent work or independent partnerships, in addition to funded projects.

To support these activities, an appropriate allocation of resources, such as computational infrastructure, development support, and data management capabilities, must be provided. This coordination is essential to maintain a balanced and sustainable infrastructure while enabling innovation and experimentation.

The results generated during the research phase may subsequently be evaluated by the Production unit, which can decide whether to adopt and further develop them. When appropriate, these outputs can be refined, stabilized, and fully integrated into the broader OSI system, transforming experimental work into operational services or databases. The components to be integrated into the OSI production unit will produce the services, data and datasets that the OSI should maintain and provide. Part of these services and databases will be exposed to the public and to potential users. It is important to emphasize that resource allocation and computational tasks should be managed through an orchestration mechanism that maintains system balance and ensures the correct operation of all OSI units. 

As outlined above, this work focuses on the operational, computational and technical aspects of the infrastructure. Accordingly, the discussion is limited to components directly involved in the IT and computational system, namely research workflows, production environments, and the supporting data, code, databases and services produced. The next subsection outlines the specific principles and key features that the OSI computational block must incorporate in order to comply with an appropriate definition of an OSI.

\subsection{OSI computational features}
Following the POSI previously introduced, which form the guiding pivots of this work, throughout our approach we outlined a systematic process, which is conceptualized by seven features to be taken into consideration and operationally integrated when defining the computational block of an OSI. These features are not derived from an existing formal framework; rather, they result from our own mapping of POSI principles to the corresponding technical requirements needed to implement them in practice. In this sense, they represent an interpretative and operational translation of POSI into concrete infrastructural concerns, grounded in implementation experience. These features are:
\begin{itemize}
    
    \item \textbf{[F1] Separation of operational domains} – to ensure proper division between special-purpose Research, Production, and Service environments (data and applications), thus separating experiments from steady operational workloads;

    \item \textbf{[F2] Service replication mechanisms} – to improve reliability, scalability, and resilience, ensuring high availability of critical services;

    \item \textbf{[F3] Centralized infrastructure orchestration} – for effective provisioning, deployment, monitoring, and management of infrastructure components;
    
    \item \textbf{[F4] Modular and interoperable architecture} – standards and proper interface definitions to allow replacing or extending components without breaking up the entire system;
    
    \item \textbf{[F5] Automation and infrastructure-as-code practices} – to ensure reproducibility and efficiency of deployment and configuration management processes;
    
    \item \textbf{[F6] Scalability and elasticity} – to make available computing and storage resources that are scalable and adaptable;
    
    \item \textbf{[F7] Reproducibility and portability of workflows} - ensuring that computational workflows can be consistently re-executed in different environments and over time.

\end{itemize}

\section{Methodology}
\label{sec:method}
Our approach provides a systematic framework for designing the technical infrastructure of an OSI. It is based on the guidelines introduced in the previous section, which represent the basics for correctly defining the OSI computational block.  

In this section, we propose a layered architecture that illustrates how each layer systematically contributes to the integration of the features F1-F7 presented in the guidelines section. Throughout this process, our methodology also emphasizes the advantages and potential limitations of selecting specific technologies, ensuring a balanced and informed approach to infrastructure design.  

A modern automated computing infrastructure can be conceptualized as a stack of interacting layers, each representing a distinct level of abstraction and system management. The scheme proposed here is inspired by widely adopted cloud computing reference-architecture concepts, particularly those described by Bohn et al. \cite{b1} and the National Institute of Standards and Technology (NIST, \url{https://www.nist.gov/}), and adapts them into four operational layers: Hardware (L1), Virtualization (L2), Orchestration (L3), and Application (L4). In addition, we introduce a Meta Layer (LM) that spans the entire computational infrastructure and influences all the other layers. In other words, it acts as a supervisory and management layer for the whole infrastructure stack.  

The following subsections discuss each layer individually. For each layer, we examine its specific role, corresponding strategic actuation, and the OSI features that it guarantees through its integration and procedures. We also determine whether each layer is mandatory or optional following its contribution in a correct definition of an OSI. Table~\ref{tab:layers} summarises the layers along with their definitions, procedures, supported OSI features, and whether their inclusion is mandatory to initiatean OSI.

\begin{table*}[t]
\caption{The table summarizes the architectural layers composing the computational block of an Open Scholarly Infrastructure (OSI), together with their role, the OSI features they guarantee, representative implementation technologies, and whether the layer is considered mandatory in the design of an OSI computational block.}
\label{tab:layers}
\setlength{\tabcolsep}{3pt}
\renewcommand{\arraystretch}{1.05}
\begin{tabular}{|p{45pt}|p{130pt}|p{140pt}|p{125pt}|p{46pt}|}
\hline
Layer &
Definition &
OSI features &
Illustrative / implementation examples &
Mandatory? \\
\hline

(L1)\par Hardware &
Provides the tangible hardware resources on which the infrastructure is built. &
-- &
Bare-metal servers &
Yes \\
\hline

(L2)\par Virtualization &
Abstracts the hardware resources into isolated and manageable execution environments. It aggregates at the node level: one or more physical hosts into multiple virtual machines. &
(F1) Separation of operational domains.\par
(F4) Modular and interoperable architecture.\par
(F6) Scalability and elasticity. &
Proxmox Virtual Environment;\par
Harvester &
No \\
\hline

(L3)\par Orchestration &
Automates the deployment, coordination, and management of services and workloads. &
(F1) Separation of operational domains.\par
(F2) Service replication mechanisms.\par
(F3) Centralized infrastructure orchestration.\par
(F4) Modular and interoperable architecture.\par
(F5) Automation and infrastructure-as-code practices.\par
(F6) Scalability and elasticity. &
Kubernetes &
Yes \\
\hline

(L4)\par Application &
Delivers the software services and functionalities directly used by end users. &
(F4) Modular and interoperable architecture. &
Containerized applications using Docker &
Yes \\
\hline

(LM)\par Meta &
Defines meta-configurations, workflows, and strategic principles governing the computational infrastructure. &
(F5) Automation and infrastructure-as-code practices.\par
(F7) Reproducibility and portability of workflows. &
Using OpenTofu to manage infrastructure declaratively and reproducibly &
No \\
\hline

\end{tabular}
\end{table*}

\subsection{Hardware Layer (L1)}
The Hardware Layer provides the underlying computational capacity, including processing power, storage, and networking resources. At this layer, core, sensitive, or latency-critical workloads are typically hosted on local servers and storage systems.  

On this layer, it would be reasonable to limit the use of closed building blocks of vendors’ software and hardware solutions. In other words, using commodity servers with standard x86/ARM processors and, when possible, open-source hardware would be more appropriate to avoid relying on hardware that might not be freely available to the community, thus violating the POSI. Furthermore, the local control of the hardware will be critical for security and cost efficiency in the long term. Digital sovereignty is a key aspect for this layer. Owning the server infrastructure guarantees full control over the stored data, safeguards it from external influence by foreign corporations and ensures that access is not constrained by commercial interests. This issue becomes especially critical in cloud-based deployments, where dependence on commercial providers may expose the OSI to changes in pricing models, contractual conditions, or service availability. Such developments could undermine the long-term sustainability of the infrastructure, making it necessary to adapt governance strategies and reallocate financial resources to address these evolving risks (cloud-based solutions will be discussed in Section~\ref{sec:conclusions}).

\subsection{Virtualization Layer (L2)}
The Virtualization Layer establishes executable operating environments built upon the Hardware Layer resources. Virtual machines abstract the underlying hardware into isolated environments, enabling multiple independent systems to operate on the same physical infrastructure.  

The purpose of this layer is to avoid deploying applications directly on hardware. By introducing this abstraction, it provides workload isolation, simplifies scaling, improves reproducibility across environments, and allows for more efficient utilization of computing resources \cite{b14}. Within this environment, an operating system is installed, runtime dependencies are configured, and networking is set up to enable communication with other systems and users. This layer, if integrated into the OSI, forms the first fully usable computing environment where software execution is possible. It can be deployed using established virtualization platforms such as Proxmox Virtual Environment (\url{https://www.proxmox.com/en/})\cite{b21} and Harvester (\url{https://harvesterhci.io/}). Deployment typically involves:
\begin{itemize}
\item installing the hypervisor on physical hosts;
\item creating virtual machine templates (typically using pre-configured OS images);
\item configuring storage (e.g., NFS, Ceph) and networking (e.g., VLANs, bridges);
\item setting resource limits (CPU, RAM, disk) and security policies.
\end{itemize}

These configurations integrate seamlessly with upper layers for orchestration and resource management. While integrating this layer into an OSI abstracts and efficiently manages many aspects of infrastructure operation, it is not mandatory. One may opt for a bare metal strategy (deploying directly on physical hardware) without virtualization. This approach, though less manageable, eliminates certain OSI features that virtualization would guarantee, such as scalability and separation of operational domains. These features could potentially be addressed at higher layers (in the Orchestration Layer – L3), but with reduced effectiveness \cite{b15}.  

At this layer, we might already decide to separate the resources dedicated to the Research and Production blocks introduced in  Figure~\ref{fig1}. Ideally, here we create VMs to be used in either the research or production blocks, thus addressing the need for a separation between the operational domains (F1). Having a tool (such as Proxmox) to handle these operations gives us a modular and configurable way to modify/resize such VMs according to the needs (F4), which makes the infrastructure much more feasible and scalable for future needs (F6).

\subsection{Orchestration Layer (L3)}
The Orchestration Layer combines either multiple bare metal machines (in case L2 is not adopted) or virtual machines into a cohesive, scalable resource pool. The cluster serves as a resource-pooling level by bringing together many servers and/or virtual machines into a single system. By combining processing power, memory, and storage, the cluster produces a system that can perform much better compared to any single machine. 

Clusters are used in many contexts. Their benefits are: (a) enabling horizontal scaling by simply adding more nodes, (b) distributing workloads to prevent bottlenecks, (c) providing fault tolerance so that failures in some nodes do not bring down the entire system, and (d) ensuring high availability to keep services running continuously \cite{b22}.

Different tools might be used in this layer, analysed and compared for different environments (\cite{b23,b24}). The most common and widely adopted solution is based on Kubernetes (K8s, \url{https://kubernetes.io/}) \cite{b16}, a container orchestration platform that manages pods across clusters with auto-scaling, self-healing, and service discovery. We adopt Kubernetes as the core orchestration platform in our methodology, considering its widespread adoption as the de facto standard that strengthens the reproducibility of the underlying infrastructure, which is a key requirement for OSIs \cite{b25}.  

The integration of L3 is essential for creating a functional infrastructure that efficiently separates operational units, i.e.  Research, Production, and Databases/Applications (F1). Even without a Virtualization Layer (L2), this stage requires mechanisms for computational separation, workload isolation, and scalability to ensure reliable operation across diverse workloads. Also, orchestrators handle the core tasks of deciding where applications run (F3), how many replicas exist (F2), how services communicate, and how failures are handled. They continuously ensure that the system matches a defined desired state, maintaining reliability, consistency, and efficiency.  

It’s worth mentioning that one may decide to use a different Kubernetes distribution \cite{b26}. Tools such as Rancher (\url{https://www.rancher.com/}) provide an additional layer of control over Kubernetes, with a different distribution called RKE, making it easier to deploy, monitor, and manage multiple orchestrators and clusters from a centralized interface. In essence, Rancher represents a manager for Kubernetes, which allows it to automate complex operations while retaining visibility and governance over the entire infrastructure. It effectively acts as the brain of the cluster, coordinating resources, automating decisions, and keeping workloads running smoothly, even as the environment grows or changes.

The Orchestration Layer is mandatory in order to enable centralized orchestration through unified cluster management and a continuous monitoring system of the services and replications that are created. These capabilities create a robust foundation for scalable, observable, and interoperable OSIs (F6). Furthermore, Kubernetes uses YAML configuration files to implement Infrastructure as Code (IaC) practices (F5), enabling automated infrastructure provisioning and ensuring the reproducibility of the deployed services. IaC approaches can improve the reproducibility and portability of infrastructure configurations by representing infrastructure declaratively and managing it through version controlled code \cite{b11}.

\subsection{Application Layer (L4)}
The Application Layer operates at the top of the stack, leveraging automated scaling and reliable execution from lower layers. Represents the final stage that transforms the entire infrastructure stack into tangible user outcomes, delivering scalable research services, reproducible analyses, and collaborative platforms while upholding maximum openness and portability. Without it, essential services (databases, applications) cannot be deployed, executed, or exposed to OSI users. The deployed applications rely on underlying layers but do not manage them directly: the Orchestration Layer (L3) is responsible for these aspects.

Some key points to consider here are:
\begin{itemize}
    \item \textbf{containerization} – applications must be packaged as Docker containers or similar OSI-compliant images to ensure portability, consistency, and orchestration compatibility;
    \item \textbf{open source} – maximizing the use of open-source software to ensure transparency, community support, and OSI compliance;
    \item \textbf{core value delivery} – implementing and exposing functionality via REST APIs, and processing scientific data workflows.
\end{itemize}  

A synchronized approach with containerization and version control of the service to containarize (for instance using Git) fosters a modular and interoperable behaviour (F4) of the entire application layer. 

\subsection{Meta Layer (LM)}
The Meta Layer operates across the entire infrastructure stack and enforces cross-cutting principles to ensure the infrastructure remains automated, governed, and reproducible. In particular, this layer supports automation, infrastructure-as-code practices, data governance and access control mechanisms (F5). These procedures enhance the reproducibility and portability of workflows used to define the infrastructure and specific services in it (F7). In a way, the Meta Layer does not correspond to a single operational environment but rather to a transversal control framework that governs how the entire infrastructure, with the layers L1 to L4, is configured, accessed, reproduced, and maintained over time.  

From an implementation perspective, this layer can be realized through technologies such as OpenTofu (\url{https://opentofu.org/}) to automate configuration and deployment and Git-based version control platforms, such as GitHub or GitLab, to track infrastructure and workflow definitions.  

This layer differs from the ones previously introduced, as it represents an extension proposed in this work rather than a component derived from the architectures of Bohn et al. \cite{b1} or NIST. Its purpose is to provide a meta-level perspective on the entire infrastructure, addressing cross-cutting aspects that support the definition, management, and governance of all the other layers composing an OSI. 

At present, we do not consider this layer mandatory, since some of its functionalities, such as Infrastructure-as-Code practices, can also be implemented within the Orchestration Layer (L3). Nevertheless, we argue that explicitly incorporating this layer promotes a more structured management of the infrastructure and further enhances its reproducibility (a more detailed discussion of these aspects is provided in Section~\ref{sec:conclusions}).

\section{An OSI case study: OpenCitations}
\label{sec:oc}

OpenCitations is an open science infrastructure organization for open scholarship. The work of OpenCitations is dedicated to the publication of open citation data and bibliographic metadata as Linked Open Data using Semantic Web technologies \cite{b6}.  

In January 2021, OpenCitations conducted a comprehensive self-assessment of its compliance with POSI requirements, which can be found at the official dedicated blog post \cite{b13}. The POSI that ground the proposed methodology makes OpenCitations an ideal OSI use case for illustrating the definition we have outlined through our methodology. OpenCitations demonstrates full compliance with the cornerstone principles we have outlined. Also, when focusing on the computational block of the infrastructure, as detailed in the compliance blog post \cite{b13}, OpenCitations meets essential requirements that we have mentioned previously:
\begin{itemize}
    \item \textbf{open source} – software is released under open source licenses;
    \item \textbf{open data} – datasets are available under CC0 waiver;
    \item \textbf{open access} – several data access options, such as REST APIs, SPARQL endpoints, web interfaces, and bulk downloads;
    \item \textbf{replicability} – infrastructure design enables straightforward replication by others.
\end{itemize}

Examining the practical implementations adopted by OpenCitations illustrates how the theoretically designed strategy has been concretely formalized in practice. Also, a premature process was first theoretically proposed with a previous work done by the OpenCitations team \cite{b3}.  

In this section, we first examine the computational block of OpenCitations using the layer-based architecture introduced in our methodology, highlighting how the identified OSI features are implemented and guaranteed. Then we present a set of practical scenarios showing how the infrastructure addresses typical operational tasks and real-world situations. 

\subsection{Computational block}
\begin{figure*}[t]
    \centering
    \includegraphics[width=\textwidth]{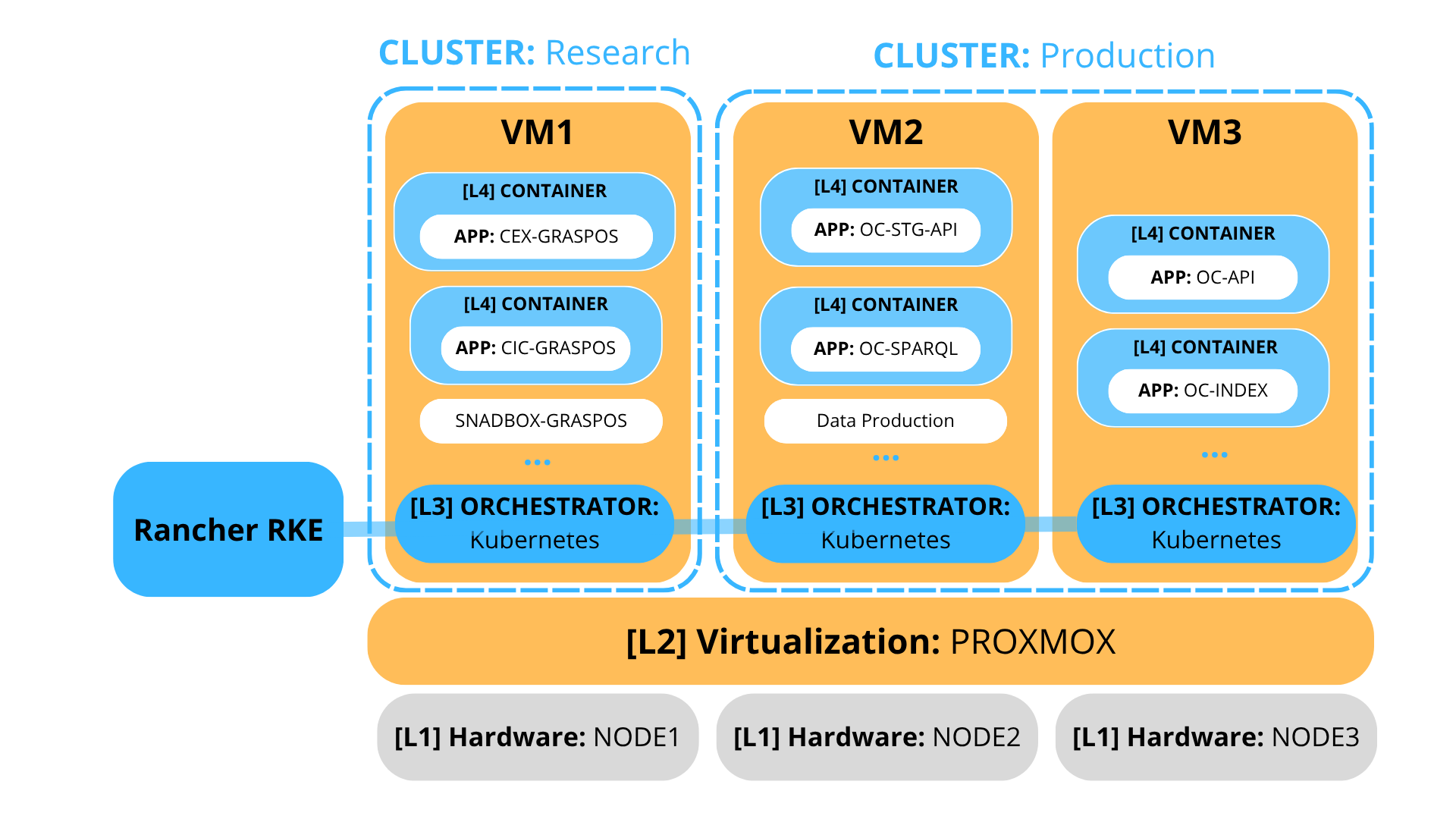}
    \caption{OpenCitations’ computational block architecture scheme. The figure illustrates all layers, from the Hardware Layer formed by 3 on-site nodes (gray) to the Virtualization Layer, including the virtual machines built using Proxmox (yellow) and the Orchestration Layer (blue) with the two clusters: implemented through Kubernetes and managed via Rancher. Regarding L4, examples of containerized applications are also shown (blue boxes with white border).}
    \label{fig2}
\end{figure*}

In this subsection, we present a concrete application of our methodology through the presentation of how the OpenCitations computational block has been defined. This case study shows how those principles can be applied in practice to a real infrastructure. Through this example, it becomes possible to observe how the different layers are implemented, how they interact with one another, and how they collectively guarantee the OSI features identified in the methodology. In this sense, the use case serves both as a validation of the proposed framework and as a practical reference for the design and assessment of comparable infrastructures. Figure~\ref{fig2} summarizes the overall computational block of the OpenCitation infrastructure architecture we are going to discuss in detail in the following subsections. All the layers we have presented in the methodology are mapped to the OpenCitations infrastructure scheme in Figure~\ref{fig2}; the scheme also includes the actual tools used and examples of the products from each layer. These parts will be addressed and become clearer in the next subsections.

\subsubsection{Hardware Layer (L1)}
The Hardware Layer is entirely based on on-premises hardware and does not rely on cloud resources. The infrastructure consists of multiple physical servers distributed across different geographic locations. Three main nodes (NODE1, NODE2 and NODE3) are a data center that provides the core computational capacity of the infrastructure. A fourth node (not included in Figure~\ref{fig2}), located at a different site, offers lower computational power and can support secondary or auxiliary functions. In addition, a QNAP NAS system located in a different geographic location of NODE1, NODE2, and NODE3 is used for long-term backup purposes, contributing to data preservation and infrastructure resilience. This distributed physical configuration provides the hardware foundation on which the higher layers are built. The computational resources of all 3 nodes will be taken into account in the upper layers, starting from the virtualization layer.

\subsubsection{Virtualization Layer (L2)}
The Virtualization Layer is implemented through Proxmox Virtual Environment, which is used to create and manage multiple virtual machines on top of the physical infrastructure. The three nodes are abstracted as separate, isolated execution environments, each with its own dedicated share of CPU, memory, and storage resources. For an illustrative reason, the overview in Figure~\ref{fig2} includes 3 VMs (VM1-VM3), yet the actual number is slightly different and is prone to changes if needed; for instance, a new VM could be initialized if requested by a specific research project which needs dedicated computational power to run some processes. Having different VMs allows us to establish a solid foundation for the cluster generation.  

Proxmox was chosen over other hypervisors for several practical reasons. It is an open-source solution based on KVM and LXC, which aligns with the infrastructure preference for free and open-source software, while still providing enterprise-grade features such as live migration, snapshot management, and a built-in web interface for user-friendly administration \cite{b27}. 

Each VM is configured as a self-contained block, with its own operating system (mostly based on Debian-based GNU/Linux distribution) and a network interface. This separation serves a dual purpose. On the one hand, it provides fault isolation: a misconfiguration or failure within one VM does not cascade to the others. On the other hand, it allows each block to be independently updated, rebooted, or reconfigured without disrupting the rest of the infrastructure. In this configuration, Proxmox forms the first operational environment on which the upper layers are built. The VMs, taken together, will then serve as the node pool that Kubernetes will consume and orchestrate into the corresponding clusters.

\subsubsection{Orchestration Layer (L3)}
At this stage, the computational resources distributed across the different virtual machines are no longer treated as isolated instances, but are instead aggregated into clusters. In particular, some VMs will be dedicated to the Kubernetes control plane/etcd, and the other VMs will represent the Kubernetes workers of the Production and Research clusters.  

The orchestration done through Kubernetes, enables the automated deployment, coordination, scaling, and management of containerized services across the aggregated infrastructure. Kubernetes does not simply provide a shared pool of resources, but actively governs how workloads are scheduled, replicated, monitored, and maintained across the available nodes. This makes it possible to ensure service continuity, improve operational efficiency, and support dynamic adaptation to changing computational demands. In particular, each service is deployed as a set of replicated containers that Kubernetes distributes across the available nodes according to resource constraints and affinity rules. If a container becomes unresponsive, health checks automatically detect the failure and trigger a restart without manual intervention, while rolling updates allow new versions of a service to be deployed gradually, replacing containers one at a time so that some instances remain available during the process. Resource boundaries for CPU and memory are enforced at the container level, preventing any single service from consuming more than its designated share and protecting the overall stability of the infrastructure. Kubernetes also handles internal service discovery and load balancing transparently: when one service needs to communicate with another, it does so via stable internal addresses that Kubernetes maintains regardless of where the underlying containers happen to be running at any given moment.  

To simplify interaction with the Kubernetes environment, OpenCitations uses Rancher as a management layer on top of it. Rancher provides a more accessible and centralized interface for controlling clusters, deploying services, and supervising workloads, thereby reducing the operational complexity of directly managing Kubernetes. In this way, the orchestration layer combines the robustness of Kubernetes with the usability benefits of Rancher, offering a more manageable and effective environment for operating the infrastructure. Rancher handles a different Kubernetes distribution called RKE. Compared with the most commonly used distributions, such as K3S, RKE is highly suitable for multi-cluster environments and is backed by SUSE’s strong team (\url{https://documentation.suse.com/cloudnative/rke2/}).

\subsubsection{Application Layer (L4)}
The Application Layer comprises the services, databases, and software components that are effectively deployed and executed on top of the orchestrated infrastructure. In OpenCitations, this layer includes production services and databases, staging services for testing and validation, temporary processes that support production workflows, and experimental applications or datasets developed by the different research units. As such, it represents the layer where the infrastructure becomes operationally meaningful, since it hosts the actual tools and resources that support both service provision and research activities.  

This layer is also strongly aligned with open-source strategies and reproducible deployment practices. The codebases and applications running within the infrastructure are made available through the organization repositories in platforms such as GitHub (\url{https://github.com/opencitations}) and Zenodo (\url{https://zenodo.org/communities/opencitations}), ensuring transparency, accessibility, and long-term preservation. Moreover, the applications are containerized through Docker, which plays a key role in packaging software and its dependencies into portable and reproducible units. This facilitates deployment in Kubernetes, simplifies maintenance across environments, and supports a more consistent and scalable management of both stable and experimental services.  

It is important to note that, as illustrated in Figure~\ref{fig2}, the applications are not always containerized. For instance, in VM2, the box “Data Production” is code that runs directly on the virtual machine rather than in a specific Docker container.  

\subsubsection{Meta Layer (LM)}
To foster the reproducibility of the applications and services deployed and orchestrated via Kubernetes all the configurations governed by the YAML files are publically available via Github repository (\url{https://github.com/opencitations/infrastructure}). The repository includes manifests regarding all the applications deployed, Helm configurations, and deployment automation required to orchestrate, manage, and scale the OpenCitations services, enabling reproducibility, maintainability across the entire infrastructure.  

The idea is to create a Git repository release for each infrastructure version that should be documented or referenced. These releases are archived through services such as Zenodo, making each version citable and easier to preserve, reference, and reproduce over time. The current status of the infrastructure is based on the v1.0.5 release \cite{b28}.  
Although this approach differs from the workflows supported by dedicated Infrastructure-as-Code tools such as OpenTofu (discussed in Section~\ref{sec:method}), it reflects the strategy currently adopted by OpenCitations for managing its infrastructure as code. At present, the main limitation to the things discussed in our methodology is the lack of a unified, ready-to-use workflow that brings together all the individual configurations and clearly demonstrates how they integrate into a complete deployment process.

\subsection{Operational Scenario: the GraspOS project}
To illustrate how the computational block of OpenCitations operates according to the architectural model already discussed, this section describes the practical scenarios that occured during the Research-Production-Services cycle that OpenCitations faced, and explains how they are handled in light of the technologies and architectural principles previously introduced.  
To show such behaviour, we illustrate how the infrastructure handled the operational procedures to be done for the GraspOS project (\url{https://graspos.eu/}) – ended in December 2025 and funded by the European Commission. The GraspOS project objective was to build and operate a data infrastructure to foster a responsible research assessment system that embeds Open Science (OS) practices and accelerates its adoption in Europe. OpenCitations (through its legal correspondence, the University of Bologna) was a partner of the project. 

From a technical point of view, which is the part directly related to what we have discussed so far, the outcomes of the GraspOS project have been focused on the production of two main tools: the Citation Extraction Service (CEX, \url{https://github.com/opencitations/cec/tree/main/extractor}), a tool used to work on the extraction of reference from PDFs \cite{b29}, and the Citation Intent Classifier (CIC, \url{https://github.com/opencitations/cec/tree/main/classifier}), a tool used for the classification of the citation intent following the in-text citation context analysis \cite{b30}. 

\begin{figure*}[t]
    \centering
    \includegraphics[width=\textwidth]{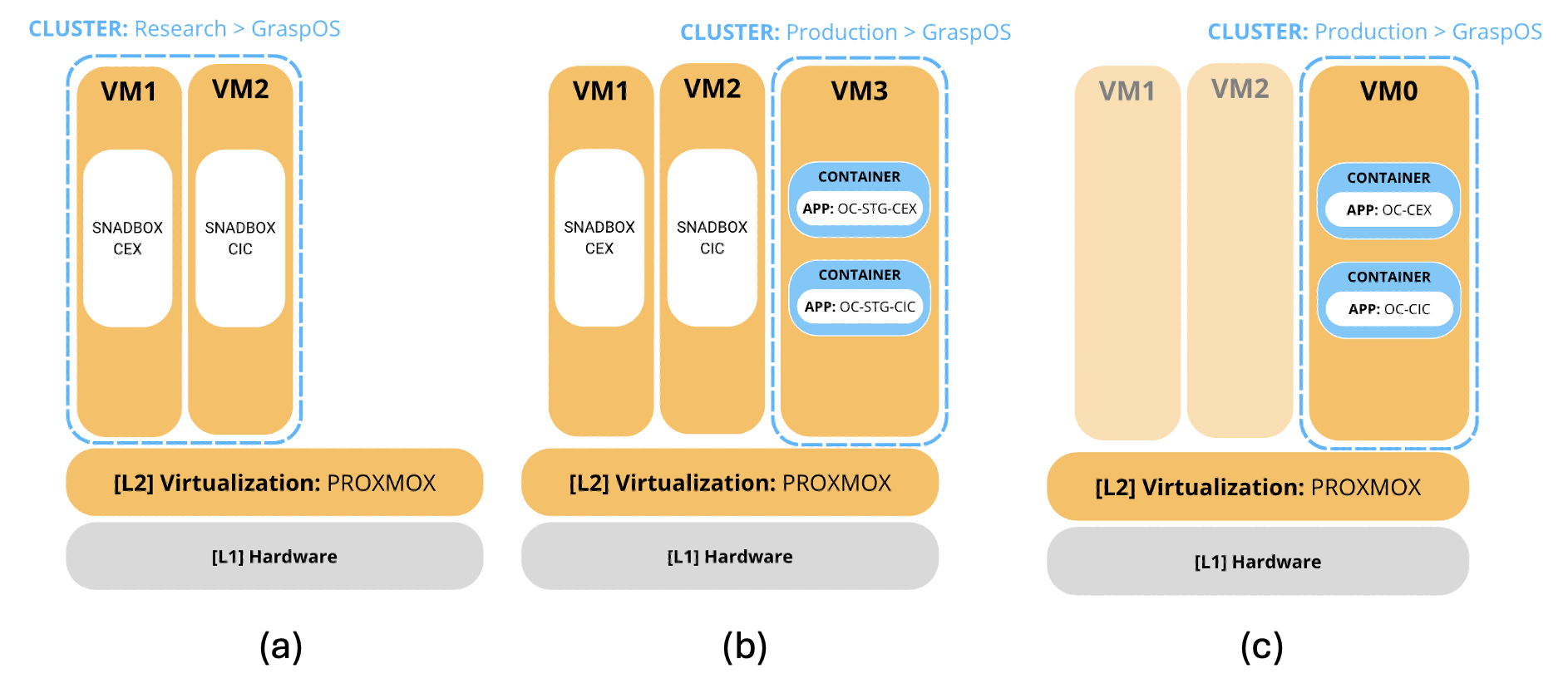}
    \caption{The computational cycle: (a) Research, (b) Production, (c) Services, for the case of the GraspOS project handled within the OpenCitations infrastructure.\label{fig3}}
\end{figure*}

In this section, we present how  the OpenCitations team conducted its work on the operational side. This case is analysed by considering the three stages of the Research-Production-Services cycle, which are treated in the following subsections with the help of the graphical illustration in Figure~\ref{fig3}, showing how the infrastructure handled the 3 stages.

\subsubsection{Research}
Using Proxmox, two dedicated VMs have been created under the “Research” cluster, each with exclusive resources. Researchers involved in the project have been granted root access to the VMs (no access to critical procedures of the infrastructure), which enabled them to freely use and develop their software and tools. At this stage, the Docker containerization of the applications was not needed. This part is illustrated in Figure~\ref{fig3} (a): VM1 and VM2 contains the two sandboxes “SANDBOX-CEX” and “SANDBOX-CIC” which represents the part of the VM running the staging code of the project without being encapsulated in any Docker.

\subsubsection{Production}
Following the research stage, the outcomes are then passed to the production unit. As expected, not all code and tools developed need to be moved to production; therefore, the OpenCitations team first filters what is necessary to include, ensuring it aligns with the project's outcome expectations. Since the software needs to be adapted to the technical capabilities of the infrastructure, they first need to be containerized in Docker, which requires refactoring parts of the code. Dedicated Dockers (\url{https://hub.docker.com/r/opencitations/oc\_cec\_extractor}, \url{https://hub.docker.com/r/opencitations/oc\_cec\_classifier}) and a GitHub repository (\url{https://github.com/opencitations/cec}) have been created within the GitHub OpenCitations organization space. Then, before moving the tools to an ongoing status, fully usable by the OpenCitations users, a test phase regarding the infrastructure compliance and performance have been performed. In Figure~\ref{fig3} (b), we have represented this part with Dockers "OC-STG-CEX" and "OC-STG-CIC", included into VM3 (part of the "Production" cluster). These Docker images will serve as the first staging version for testing.  

\subsubsection{Services}
The applications in staging are tested both functionally, to verify that outputs and behavior match expectations, and in terms of performance. Performance testing covers the single service being used as well as its overall impact on the infrastructure. Once this phase is complete and the services behave as expected, the application containers are promoted to deployment status. Deployed services may remain online for varying periods. In the GraspOS case, the two services – \textit{OC\_CEX} and \textit{OC\_CIC} – were deployed and remained running within the OpenCitations infrastructure for the entire duration of the project, in the VM0: a VM used for stable running services in the "production" cluster (see Figure~\ref{fig3} (c)).

\section{Conclusions and final considerations}
\label{sec:conclusions}
This paper proposed a layered methodology for defining the technical and computational block of an OSI, aimed at translating broad OSI principles into a concrete architectural framework. Starting from the POSI and a set of core computational features, the study identified the main technical requirements that an OSI should satisfy in order to remain open, sustainable, interoperable, reproducible, and operationally reliable.  

The proposed framework organizes the computational block into four main layers: Hardware (L1), Virtualization (L2), Orchestration (L3), and Application (L4), complemented by a transversal Meta Layer (LM). This layered view makes it possible to clarify the role of each infrastructural level, to distinguish mandatory from optional components, and to show how different technical features can be systematically integrated into the overall definition of an OSI.  

The OpenCitations case study demonstrated how this framework can be instantiated in practice through a real on-premises infrastructure. By analysing its computational architecture and the operational scenario of the GraspOS project, the article shows how the interaction between research environments, production workflows, orchestration mechanisms, and deployed services can be managed coherently within a layered OSI model. This confirms that the proposed approach is not only theoretically consistent but also applicable to a functioning open scholarly infrastructure.  

We classify the Meta Layer as optional because its functionality can, to some extent, be subsumed by the orchestration layer. Nevertheless, we argue that explicitly modeling it as a distinct layer provides substantial benefits. This choice reflects the fact that some of its functionalities, such as Infrastructure-as-Code practices, can also be implemented within the orchestration layer via Kubernetes YAML configuration files. However, we argue that a truly reproducible infrastructure should extend beyond service orchestration and encompass the entire layered architecture presented in this work. By explicitly introducing a dedicated Meta Layer, the infrastructure can be managed through a unified and reproducible workflow that captures not only deployment procedures but also the relationships and dependencies among all architectural layers. The adoption of such an approach requires a mature governance model together with organizational processes capable of maintaining the infrastructure specifications, automation workflows, and resource allocation over time.  

Another important design consideration concerns the adoption of cloud-based solutions within the Hardware Layer. In many cloud environments, hardware resources are provisioned together with provider-managed virtualization and orchestration mechanisms. While this approach offers significant advantages in terms of scalability and operational simplicity, it also introduces dependencies that may conflict with the long-term sustainability objectives of an OSI. In particular, reliance on commercial cloud providers may expose the infrastructure to changes in pricing models, contractual conditions, service availability, or hardware offerings, all of which are beyond the direct control of the infrastructure’s technical management. Such dependencies may ultimately affect the reproducibility, governance, and long-term preservation of the infrastructure. For these reasons, although we do not exclude the use of cloud technologies, we recommend in-house hardware deployment as the preferred approach for the core infrastructure. To avoid relying solely on commercial cloud service providers, one possible solution would be to leverage institutional platforms such as EOSC or CERN, particularly if this need becomes more prominent among infrastructures aiming to implement FAIR- and POSI-compliant solutions. Also, cloud resources can instead be effectively employed to provide burst capacity, disaster recovery (DR) capabilities, or additional computational resources that complement, rather than replace, the infrastructure required to operate the essential services of an OSI.

\section{Declarations}
\subsection{Author's contribution statements}
\begin{itemize}
    \item \textbf{Ivan Heibi}: Conceptualization, Investigation, Methodology, Resources, Software, Validation, Writing – original draft, Writing – review \& editing.
    \item \textbf{Mario Petrella}: Conceptualization, Investigation, Methodology, Resources, Software, Writing – review \& editing.
    \item \textbf{Angelo Di Iorio}: Methodology, Supervision, Writing – review \& editing.
    \item \textbf{Silvio Peroni}: Conceptualization, Funding acquisition, Methodology, Resources, Software, Supervision, Writing – original draft, Writing – review \& editing.
\end{itemize}

\begin{IEEEbiographynophoto}{Ivan Heibi} (https://orcid.org/0000- 0001- 5366- 5194) received the Ph.D. degree in computer science from the University of Bologna, where his doctoral research focused on applying Semantic Web technologies to the Science of Science domain, particularly in the Arts and Humanities. He is currently an Assistant Professor at the University of Bologna, Department of Classical Philology and Italian Studies, working primarily on the PE5 – CHANGES project and involved in Virtual Technologies for Museums and Art Collections. He has contributed to the development and management of the OpenCitations infrastructure, where he is the Chief Technology Officer. His publications include articles on open citation data and semantic artefacts in international venues. His research interests include Semantic Web technologies, Information Science, bibliometrics, data visualisation, and automatic document analysis. Prof. Heibi is a member of the Research Centre for Open Scholarly Metadata and the Digital Humanities Advanced Research Centre.
\end{IEEEbiographynophoto}

\begin{IEEEbiographynophoto}{Mario Petrella} graduated in DAMS (Drama, Art and Music Studies) at the University of Bologna. Over the years, he completed various IT courses and earned several certifications. In 2024, he joined the OpenCitations team, where he is responsible for IT infrastructure duties and Infrastructure as Code development. He currently holds an administrative position within the FICLIT Department (Department of Classical Philology and Italian Studies) at the University of Bologna, where he is responsible for OpenCitations’ IT infrastructure.
\end{IEEEbiographynophoto}

\begin{IEEEbiographynophoto}{Angelo Di Iorio}(https://orcid.org/0000-0002-6893-7452) is Associate Professor at the Department of Computer Science and Engineering (DISI) of the University of Bologna. He holds a Ph.D. in Computer Science from the University of Bologna. His research interests also include: semantic publishing, versioning and diff-ing techniques, collaborative editing, Web technologies, formatting and pagination. Prof. Di Iorio is a member of the Research Centre for Open Scholarly Metadata and the Digital Humanities Advanced Research Centre.
\end{IEEEbiographynophoto}

\begin{IEEEbiographynophoto}{Silvio Peroni}(https://orcid.org/0000-0003-0530-4305) received the Ph.D. degree in computer science from the University of Bologna. He is currently an Associate Professor at the University of Bologna, where he works in the Department of Classical Philology and Italian Studies. He is the Director of the Research Centre for Open Scholarly Metadata and OpenCitations and a main developer of the SPAR (Semantic Publishing and Referencing) Ontologies. His publications include the book Semantic Web Technologies and Legal Scholarly Publishing (Springer, 2014) and articles in leading international journals. His research interests include Semantic Web technologies, ontology modelling, scholarly communication, bibliometrics, and open science infrastructures.
\end{IEEEbiographynophoto}

\newpage

\EOD

\end{document}